\documentclass[prl,amsmath,amssymb,preprint,superscriptaddress]{revtex4}

\usepackage{amsmath}    
\usepackage{amsfonts}
\usepackage{amssymb}
\usepackage{graphicx}   
\usepackage{dcolumn}
\usepackage{bm}

\begin{document}

\title{Particles trajectories in magnetic filaments}

\author{A. Bret}
\affiliation{ETSI Industriales, Universidad de Castilla-La Mancha, 13071 Ciudad Real, Spain}
 \affiliation{Instituto de Investigaciones Energ\'{e}ticas y Aplicaciones Industriales, Campus Universitario de Ciudad Real,  13071 Ciudad Real, Spain.}

\date{\today }

\begin{abstract}
The motion of a particle in a spatially harmonic magnetic field is a basic problem involved, for example, in the mechanism of formation of a collisionless shock. In such settings, it is generally reasoned that particles entering a Weibel generated turbulence are trapped inside it, provided their Larmor radius in the peak field is smaller than the field coherence length. The goal of this work is to put this heuristic conclusion on firm ground by studying, both analytically and numerically, such motion. A toy model is analyzed, consisting of a relativistic particle entering a region of space occupied by a spatially harmonic field. The particle penetrates the magnetic structure in a direction aligned with the magnetic filaments. Although the conclusions are not trivial, the main result is confirmed.
\end{abstract}

\maketitle

\section{Introduction}
The filamentation instability, triggered when two counter-streaming plasmas collide, grows a nearly spatially harmonic magnetic field \cite{Fried1959,BretPoPReview}. When it reaches saturation, it leaves an array of magnetized filaments which start merging \cite{Milos2006,Spitkovsky2008,Chang2008,Davis2013,Bochkarev2015}. If plasma keeps flowing in the region occupied by this field, the question of the particles' trajectories in it comes into play. This issue is specially relevant to the mechanism of collisionless shocks formation. In this respect, Particle-In-Cell simulations have shown that the shock starts forming when the peak magnetic field, together with its coherence length ($1/k$), are large enough to stop the incoming flow in the region where the instability grew \cite{BretPoP2014,Bret2015ApJL}. An intuitive case, which proves correct, has been made that for this the occur, the field coherence length, i.e. the size of the magnetic filaments, must be larger than the Larmor radius of the incoming particle in the peak field \cite{lyubarsky06,BretPoP2013}.

\begin{figure}
\includegraphics[width=0.6\textwidth]{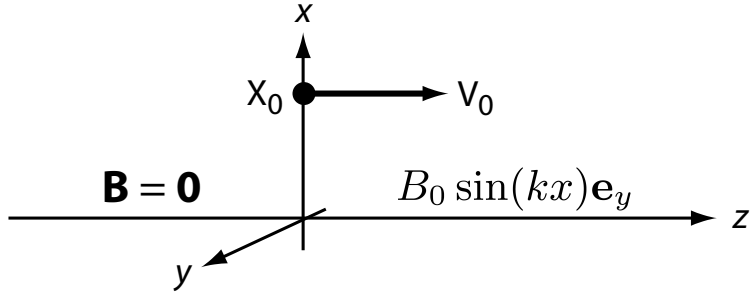}
\caption{Setup considered. The field is $0$ for $z<0$ and $B_0\sin(kx)\mathbf{e}_y$ for $z\geq 0$. Particles arrive in the field region with $\mathbf{x}(t=0)=(x_0,0,0)$ and $\dot{\mathbf{x}}(t=0)=(0,0,v_0)$.}
\label{fila}
\end{figure}

The goal of this work is to put this reasoning on firm ground, by providing a rigourous description of the particles' trajectories in a spatially harmonic field. To do so, we model the process as pictured on Figure \ref{fila}. Note that the very nature of the problem forbids us to use any slowly varying field concept, like the adiabatic invariant for example \cite{boyd}.

We consider a particle of mass $m$ and charge $q$ travelling through an inhomogeneous magnetic field. The field $\mathbf{B}$ reads $0$ for $z<0$ and $B_0\sin(kx)\mathbf{e}_y$ for $z\geq 0$. At $t=0$, the position of the particle is $\mathbf{x}(t=0)=(x_0,0,0)$ and its velocity $\dot{\mathbf{x}}(t=0)=(0,0,v_0)$, with $v_0>0$. Note that we conduct a test-particle type approach of the same kind that the one conducted in the heuristic reasoning described earlier. Collective effects could be investigated in further works. Further caveats of the model are listed in the conclusion.

Since the particle is subjected to a magnetic field only, its Lorentz factor $\gamma$ is constant. The equation of motion for $z\geq 0$ then reads,
\begin{equation}\label{motion}
m \gamma \ddot{\mathbf{x}} = q \frac{\dot{\mathbf{x}}}{c}\times \mathbf{B}.
\end{equation}
Setting,
\begin{equation}\label{variables}
\omega_B = \frac{q B_0}{\gamma m c}, ~~~~ \mathbf{x}\rightarrow \mathbf{X}/k,~~~~t \rightarrow \tau/\omega_B.
\end{equation}
Equation (\ref{motion}) reads for $Z\geq 0$,
\begin{eqnarray}\label{eq_dimless}
\ddot{X}       &=&   -\dot{Z}\sin X  \nonumber \\
\ddot{Y}       &=&   0  \nonumber \\
\ddot{Z}       &=&    \dot{X}\sin X
\end{eqnarray}
with initial conditions,
\begin{eqnarray}\label{ini:dimless}
\mathbf{X}(\tau=0)&=&(kx_0,0,0), \nonumber \\
\mathbf{\dot{X}}(\tau=0)&=&\left(0,0,\frac{kv_0}{\omega_B} \right).
\end{eqnarray}
The initial dimensionless velocity $\dot{Z}_0=kv_0/\omega_B > 0$, is simply the ratio of the Larmor radius to the filament size.

The second equation, $\ddot{Y}=0$, coupled with the initial conditions above, simply means that the motion lies in the $(X,Z)$ plane. The problem is bi-dimensional. The differential system under scrutiny is eventually,
\begin{eqnarray}
\ddot{X}       &=&   -\dot{Z}\sin X H(Z),  \label{eq_dimless1} \\
\ddot{Z}       &=&    \dot{X}\sin X H(Z),  \label{eq_dimless2}
\end{eqnarray}
where $H(Z)$ is the step function, $H(Z)=0$ for $Z<0$ and $H(Z)=1$ for $Z\geq 0$. The initial conditions are defined from (\ref{ini:dimless}) as,
\begin{eqnarray}\label{eq_dimlessCI}
X_0         &=&   kx_0, \nonumber \\
\dot{Z}_0   &=&   kv_0/\omega_B.
\end{eqnarray}

\begin{figure}
\includegraphics[width=0.6\textwidth]{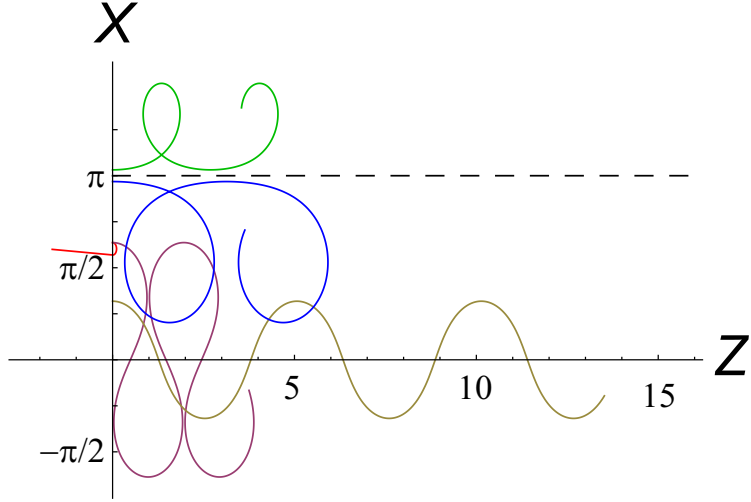}
\caption{(Color online) Numerical solutions of the system (\ref{eq_dimless},\ref{ini:dimless}). The red and the purple trajectories both have $X_0=2$. But the red one has $\dot{Z}_0=0.1$ while the purple one has $\dot{Z}_0=0.9$. The beige trajectory starts with the same velocity than the purple one, $\dot{Z}_0=0.9$, but from  $X_0=1$. The red trajectory returns to $Z<0$ while the others proceed to $Z = +\infty$. Starting from $X_0=\pi\pm 0.1$ respectively, the green and blue trajectories are bounded in the $X$ direction by the inequality (\ref{eq:bound}). The horizontal line at $X=\pi$ is never crossed (see text).}
\label{traj}
\end{figure}

Before we move to some analytical results, let us plot a few trajectories by numerically solving our system. To this extent, Eqs. (\ref{eq_dimless1},\ref{eq_dimless2},\ref{eq_dimlessCI}) have been solved using the \textbf{NDSolve} function of the \emph{Mathematica} software. Figure \ref{traj} shows how, depending on their initial conditions, some trajectories return to $Z<0$, while others proceed to $Z = +\infty$. Our goal is to analyze the ultimate evolution of the trajectories in terms of their initial conditions.

\begin{figure}
\includegraphics[width=0.9\textwidth]{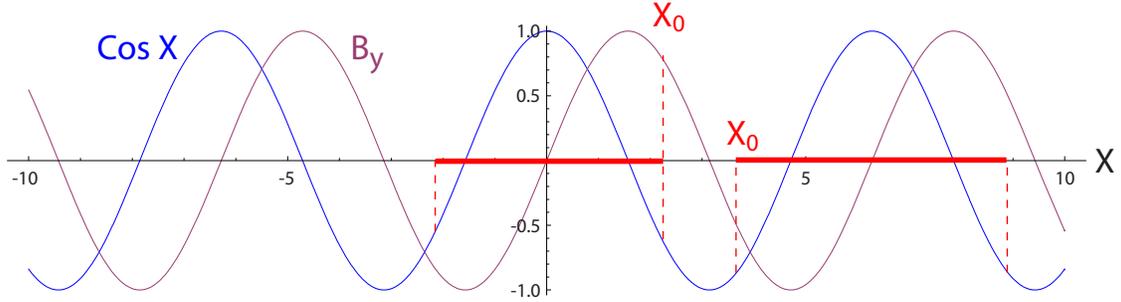}
\caption{(Color online) Trajectories are bounded by Eq. (\ref{eq:bound}) within intervals of width $<2\pi$.}
\label{bound}
\end{figure}

\section{Analytical results}
\subsection{$X$-range and -periodicity}
We here establish that the motion is always bounded in the $X$ direction, which will eventually allow to restrict our study to $X_0 \in [0,\pi]$. Integrating (\ref{eq_dimless2}) from 0 to $\tau$ gives
\begin{equation}\label{eq:int1}
\dot{Z}(\tau) - \dot{Z}_0 = \cos X_0 - \cos X.
\end{equation}
Since the velocity is constant in the magnetic field and $\dot{X}_0=0$, we know that  $\dot{Z}(\tau) \leq \dot{Z}_0$. Therefore, and as a consequence of the equation above, $\cos X_0 - \cos X \leq 0$, i.e,
\begin{equation}\label{eq:bound}
\cos X \geq \cos X_0.
\end{equation}

Depending on the sign of $\sin X_0$, the particle is initially deflected upward (if $\sin X_0<0$), or downward (if $\sin X_0>0$). Figure \ref{bound} illustrates how these constraints result in a confinement of the trajectory within intervals of width $<2\pi$. As a consequence, we can restrict our study to  $X_0 \in [-\pi,\pi]$. In addition, the system (\ref{eq_dimless1},\ref{eq_dimless2}) is clearly invariant when changing $X \rightarrow -X$, so that it is enough to investigate the interval $X_0 \in [0,\pi]$.

\subsection{Sufficient condition for $\lim_{\tau\to\infty} Z(\tau) = \infty$}
Integrating now (\ref{eq:int1}) from 0 to $\tau$ and accounting for $Z_0=0$, we find,
\begin{equation}
Z(\tau) =   \int_0^{\tau}(\dot{Z}_0 + \cos X_0 - \cos X) d\tau.
\end{equation}
From $-1\leq- \cos X\leq1$, we deduce the following inequality for the integrand,
\begin{equation}
\dot{Z}_0 + \cos X_0 -1 \leq \dot{Z}_0 + \cos X_0- \cos X \leq \dot{Z}_0 + \cos X_0 + 1.
\end{equation}
Therefore,
\begin{itemize}
   \item If $\dot{Z}_0 + \cos X_0 -1 \geq \alpha$ for any $\alpha > 0$, then $Z(\tau) > \alpha\tau$ and $\lim_{\tau\to\infty} Z(\tau) = \infty$. The particle streams through the magnetized region.
   \item If $\dot{Z}_0 + \cos X_0 + 1 \leq -\alpha$ for any $\alpha > 0$, then $Z(\tau) < -\alpha\tau$ and $\lim_{\tau\to\infty} Z(\tau) = -\infty$. The particle bounces back on the magnetized region.
 \end{itemize}
Note that in the second case, our analysis is only valid as long as $Z \geq 0$. Yet, as soon as the particle with an initial $\dot{Z}_0 > 0$ crosses back the frontier $Z = 0$ from $Z > 0$, it finds itself in a field-free region where it won't be deflected again toward a positive value of $Z$. Hence, our conclusion remains unchanged: the particle returns back to $Z = -\infty$.

The first conclusion is already a partial confirmation of the intuitive reasoning that the threshold for bouncing back is $\dot{Z}_0 \sim 1$. Averaging over the possible values of $X_0$, we find particles stream through the magnetized region if $\dot{Z}_0 > 1$, i.e. $v_0/\omega_B > k^{-1}$. This is precisely equivalent to stating that the filaments are too small compared to the Larmor radius of the incoming particle in the peak field.

The second conclusion is indeed trivial since $\dot{Z}_0 + \cos X_0 + 1 \leq 0$ eventually amounts to $\dot{Z}_0  \leq - \cos X_0 - 1 \leq 0$: if the particle is launched from $Z_0=0$ with $\dot{Z}_0  \leq 0$, it necessarily goes to $Z=-\infty$, or stay at $Z=0$ (in the case $\dot{Z}_0 = 0$).

\subsection{Linearization for $\sin X \ll 1$}
A more accurate treatment of the system (\ref{eq_dimless1},\ref{eq_dimless2}) can be performed if $\sin X_0\sim 0$. Let us start assuming that $\sin X_0\sim 0$ implies  $\sin X(\tau)\sim 0$, $\forall \tau>0$, and check the assumption afterward. If $X_0\sim 0$, we'll then write $\sin X \sim X$. If $X_0 = \pi$, we can replace $X \rightarrow X - \pi$ and analyze the system considering $\sin X \sim -X$. The system of linearized equations eventually reads,
\begin{eqnarray}
\ddot{X}       &=&   - \epsilon \dot{Z} X,  \label{eqred_linea1} \\
\ddot{Z}       &=&    \epsilon\dot{X} X,   \label{eqred_linea2}
\end{eqnarray}
with $\epsilon=1$ if $X_0\sim 0$, $\epsilon=-1$ if $X_0\sim \pi$, and initial conditions (\ref{ini:dimless}). Integrating Eq. (\ref{eqred_linea2}) gives,
\begin{equation}
\dot{Z} = \dot{Z}_0 + \frac{\epsilon}{2}X^2 -\frac{\epsilon}{2}X_0^2.
\end{equation}
Note that this equation is nothing but the linearization of Eq. (\ref{eq:int1}) for $\sin X \ll 1$. Inserting this expression of $\dot{Z}$ in Eq. (\ref{eqred_linea1}) and neglecting second order $X$ terms and superiors (including $X_0^2$), we find,
\begin{equation}
\ddot{X} +  \omega^2 X  = 0, ~~~ \omega^2 = \epsilon \dot{Z}_0 ~~ \Rightarrow ~~ X(\tau) = X_0\cosh (i\omega \tau).
\end{equation}
For $X_0\sim \pi$, $\epsilon=-1$ and  $i\omega$ is real. Here, trajectories quickly depart from the region where $\sin X \sim -X$, and linearization breaks down. The trajectory $(X=\pi,Z=\dot{Z}_0\tau)$ is unstable. But for $X_0\sim 0$, $\epsilon=1$ and  $X(\tau) = X_0\cos (\omega \tau)$: trajectories remain confined within $X\in [-X_0,X_0]$, and the linear treatment remains valid. Using then $\ddot{Z} =  \dot{X} X$ and integrating twice gives,
\begin{equation}\label{eq:Zanal}
  Z(\tau) = \left(\dot{Z}_0-\frac{X_0^2}{4}\right)\tau +\frac{\sin (2\omega\tau)}{8 \omega }X_0^2.
\end{equation}
The particle is thus repelled from the magnetized region if $\dot{Z}_0 < X_0^2/4$, a slightly more stringent condition than $\dot{Z}_0 < 1-\cos X_0\sim X_0^2/2$, near $X_0=0$.

\begin{figure}
\includegraphics[width=0.6\textwidth]{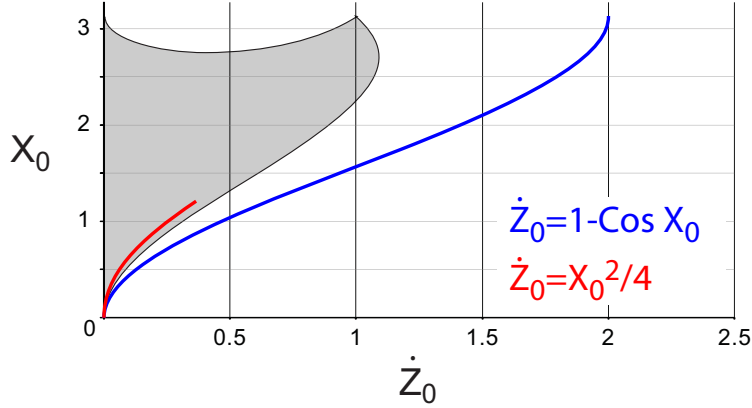}
\caption{(Color online) Whenever the initial conditions $(\dot{Z}_0,X_0)$ pertain to the shaded area, the particle bounces back.}
\label{phase}
\end{figure}

\section{Numerical results}
In order to derive, for any $X_0$, the sufficient and necessary condition for particles to bounce back against the magnetized region, we numerically explore the phase space $(\dot{Z}_0,X_0)$. For any given $X_0 \in [0,\pi]$, we scan the values of $\dot{Z}_0$ yielding the particle to bounce.

The result of this numerical exploration is summarized in Figure \ref{phase}: whenever the initial conditions $(\dot{Z}_0,X_0)$ pertain to the shaded area, the particle bounces back. The full picture can be completed by symmetry around $X_0=\pi$, and $X_0=0$.

For $X_0\sim 0$, we numerically find the bouncing threshold at $\dot{Z}_0 \sim 0.34 X_0^2$. The discrepancy with the $\dot{Z}_0 = X_0^2/4$ condition derived in the linear regime is explained as follow: some forward trajectories, quite like the purple one on Fig. \ref{traj}, come close to exit the $Z\geq 0$ region by their beginning. Even if they would go to $Z=+\infty$ in a space filled-up with the harmonic field, they leave the magnetized region early in time, simply because they wander into the field-free zone $Z<0$ during their first oscillation. Trajectories with $X_0\sim 0$ are forward (backward) for $\dot{Z}_0 > (<) X_0^2/4$. But they wander into the field free zone as soon as $\dot{Z}_0 < 0.34 X_0^2$.

This factor 0.34 can be approximated analytically in the following way. One starts linearizing Eq. (\ref{eq:Zanal}) up to the 3rd order in the time parameter $\tau$. Equaling the result to 0 gives the moment $\tau_0$ when the particle, starting from $Z=0$, will reach $Z=0$ again. The result is $\tau_0=\sqrt{6\dot{Z}_0}/\omega X_0$ \footnote{This procedure obviously requires $2\omega \tau_0\ll 1$, i.e. $\dot{Z}_0 \ll X_0^2/24$. Although this strong inequality is not fulfilled, the result is quite close to the numerical one.}. Then, if $\dot{Z}(\tau_0)< 0$, the particle will definitively escape toward the $Z<0$ region. The time derivative of Eq. (\ref{eq:Zanal}) now readily gives $\dot{Z}(\tau)$. Setting $\tau=\tau_0$ in the result and developing it up to the 3rd order in $\dot{Z}_0$ gives $\dot{Z}(\tau_0)< 0 \Leftrightarrow \dot{Z}_0 < \frac{1}{3} X_0^2$, which is reasonably close to $\dot{Z}_0 < 0.34 X_0^2$.

For $X_0\sim\pi$ the shaded area displays a ``hole'' between $\dot{Z}_0=0$ and $1$. We here deal with trajectories like the blue or the green ones on Fig. \ref{traj}. As stated earlier, the linear analysis fails here because trajectories quickly departs from $X\sim\pi$. But the numerical investigation shows that these trajectories starting from $X_0\sim\pi$ have a common point with those starting near $X_0$: because the field is zero a both locations, particles starting exactly from $X_0=\pi$ or 0 definitely stream through. When starting slightly aside from these values, it only takes them a small initial velocity to override the field and move forward.

\begin{figure}
\includegraphics[width=0.7\textwidth]{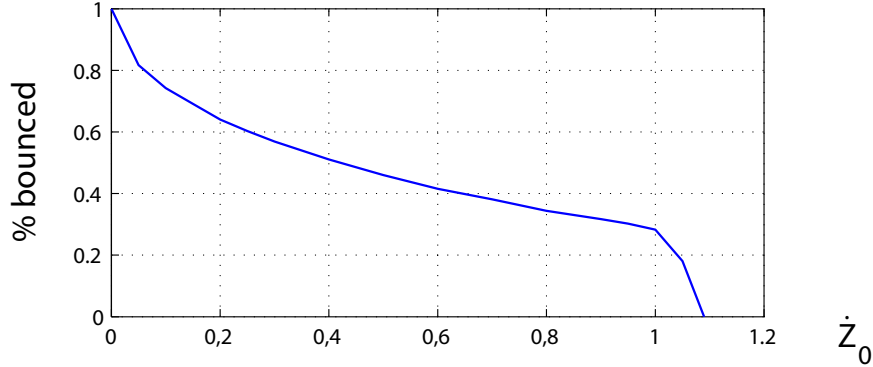}
\caption{Portion of a spatially uniform incoming beam bounced back against the magnetized region, in terms of $\dot{Z}_0$.}
\label{bounce}
\end{figure}

This gap in the bouncing region forbids to average over $X_0$ and derive an averaged bouncing, or streaming through, condition. We can instead assume an homogenous incoming beam covering all values of $X_0$, and numerically derive from Fig. \ref{phase} the portion of it which is bounced back, in terms of $\dot{Z}_0$. The result is pictured on figure \ref{bounce}. For zero initial velocity, 100\% is bounced back. The percentage then decreases until it reaches 0 for $\dot{Z}_0 \sim 1.09$. We could not find a way to approximate this number analytically. Macroscopically, we thus recover a condition for the beam to stream through the magnetized region, of the form $\dot{Z}_0 > 1$, that is, $k^{-1} < v_0/\omega_B$.

\section{Conclusion}
When dealing with more realistic settings like the ones arising from Particle-In-Cells simulations, the setup is far from being so simple. In this respect, the model should be refined in at least 3 ways.
\begin{enumerate}
  \item The filamentation instability does not grow a single mode, but a continuum of modes instead. Although the fastest growing one in the linear phase definitely governs the end picture, some smaller and larger wavelengths also grow at a slightly slower rate \cite{GremilletPoP2007}, so that the resulting field is more involved than the present single harmonic one.
  \item The transition from the field-free region to the magnetized one is not clear cut, but displays an involved gradient instead \cite{Pathak2015}. In addition, the existence of electric fields generated by charge separation has been demonstrated in this transition layer \cite{Milosavljevic2006}. Finally, still in the transition region, the longitudinal wavelength is comparable to the transverse wavelength so that the semi-infinite filaments model fails.
  \item It is assumed that $B=0$ for $z<0$. In a realistic shock configuration, this might be identified with the upstream region. However, the upstream region is not unmagnetized, but it has magnetic filaments that are roughly stationary in the fluid frame. This implies that, for a particle moving with the $\mathbf{E} \times \mathbf{B}$ drift, the force will vanish, similarly to the case of a region with zero electric and magnetic fields (as assumed in this work). However, the situation will be different for particles reflected back upstream by the shock. In the present model, they will continue towards upstream infinity, whereas in a more realistic shock they would be advected back towards the shock.
\end{enumerate}

The present model seems therefore more appropriate to the early stages of formation of a collisionless unmagnetized shock. Here, particles with $\dot{Z}_0 > 1$, i.e. $v_0/\omega_B > k^{-1}$, mostly stream through the magnetized region \cite{BretPoP2014}. But those with $v_0/\omega_B < k^{-1}$ are allowed to pass through the interface, before they are trapped in the magnetized region (instead of bouncing back). This is the onset of the collisionless shock formation. At any rate, we have checked that $v_0/\omega_B \sim k^{-1}$ is the threshold for two completely different qualitative behaviors. The incoming flow gets trapped in the Weibel generated turbulence, if the Larmor radius of the particles in the peak field is smaller than the size of the filaments.

This work was supported by grant ENE2013-45661-C2-1-P from the Ministerio de Educaci\'{o}n y Ciencia, Spain
and grant PEII-2014-008-P from the Junta de Comunidades de Castilla-La Mancha. Thanks are due to C\'{e}sar Huete Ruiz de Lira for enriching discussions.


\end{document}